\documentclass[12pt]{article}
\usepackage{blois,graphicx}

\begin{document}

\title{HE Stratosphere Event of 1975 Revisited: the Difference between the Patterns of Astroparticle Interaction and LHC Nucleus-Nucleus Collision.}
\author{O.I. Piskounova and K.A. Kotelnikov}
\address{P.N.Lebedev Physics Institute of Russian Academy of Science, Leninski prosp. 53, 119991 Moscow, Russia}
\maketitle\abstract{
The event of astroparticle collision at high energy was detected in 1975 during the balloon flight in the stratosphere. 
The data of hundred particle tracks in x-ray films have been re-analyzed in the style of LHC experiments: 
rapidity distributions of charged particles and transverse mass spectra of multi-particle production have been built.  
The comparison of multiple histograms with the expectations of the Quark-Gluon String Model (QGSM) gives us, at first sight, the conclusion that it might be the carbon-nucleus collision with the matter of atmosphere at the c.m.s. equivalent energy $\sqrt{s} \ge $ 5 TeV. 
After QGSM analysis of these scarce data, we know the following: the value of maximal rapidity of one projectile proton and
the density of particle multiplicity in the central rapidity region. Besides this, the transverse mass distributions show us how many protons are in every particular range of rapidity. In such a way, we certainly can distinguish how this astroparticle interaction is similar to or differs from the average A-A collision event at LHC. Nevertheless, the data indicate the features that cannot be associated with nucleus-nucleus collision: one particle with transverse mass 16 GeV was detected and a small nucleon population has been seen in the region of projectile fragmentation that does not correspond to the carbon nucleus collision. Both facts make us convinced that there might be a baryonic DM decay. These heavy quasi-stable baryon-antibaryon neutral states  have been suggested in the earlier paper (Piskounova O., 2018). They are to be formed at the huge gravitation pressure in giant massive objects like Black Holes. Relativistic jets are spreading the baryonic DM in space. Their collisions with ordinary matter have to give the different pattern than A-A interactions. The important difference between new form of matter and the ordinary nucleus lies in the results of collision: baryonic DM is the object, where protons and antiprotons are strongly connected, so the energy is devided between the components of this astroparticle due to the structure function of Regge type, like quarks in the proton. The lightest debris of baryonic DM particle interacts with the greatest maximal rapidity and gives the small number of nucleons in the forwarding part of spectra. Baryonic DM can also split into the pair of similar DM with lower mass giving  unusual couple of hadrons with mass near 14 GeV and heavier.
Finally, we conclude that the rsults of cosmic ray experiments on the high altitudes in the atmosphere are, on one hand, good supplements 
to the LHC measurements. On the other hand, this experiments are able to discover events of unknown  astroparticle collisions in the full kinematical range, while colliders are studying nuclear interactions only in the central rapidity region. Such experiments that detect the very first collision of astroparticles with the atmosphere should be continued, but they are prefered to be constructed with the application of up-to-date electronic methods.
}

\section{Introduction}

At least three particles were unpredictably discovered  in cosmic rays: positron - in 1929 (Dmitri Skobeltsyn) and in 1933 (Carl Andersen), muon - in 1936 (Carl D. Anderson and Seth Neadermeyer) and kaon - in 1947 (K.K.Batler and G.Rochester). Nowadays, the collider experiments are based on the concrete predictions of contemporary theories. In this case, any research seems unable to meet an observation of unpredicted new particles or phenomena.

An impressive event of HE interaction was detected long ago at the balloon flight in the stratosphere \cite{stratosphere}. 
Here we are re-analyzing this event with the methods of data presentation of LHC collider experiments \cite{root}.
The small statistics make our analysis rather challenging, as it is typical for hadroproduction measurements in cosmic ray physics. We hardly distinguish the type of produced particles as well as the signs of their charge.
Fortunately, the Model of Quark-Gluon Strings (QGSM) \cite{QGSM,myPhD} gives us the idea about how the rapidity spectra of hadroproduction looks like in the full rapidity interval of projectile fragmentation. The QGSM was developed in the early 80'th for the purpose of complementary studies of HE fixed target hadron collisions in cosmic ray physics as well as in p-p and A-A interactions at colliders. It helps us also to conclude about new phenomena, if the sings of them are existing in the available spectra both at the central rapidity area and in the forward fragmentation region.   
The region of target nucleus fragmentation is not seen due to the small surface of the detector, while the rapidity distribution in the forward fragmentation region is seen completely, so the important information about a projectile particle is to be reconstracted. As an example, rapidity spectrum has to show few protons at the energy area of three Pomeron peak, as it was predicted in QGSM \cite{kaidalov}.
The additional information was obtained from the target diagram of produced particles. The analysis of transverse coordinates of detected particles shows the transverse mass distribution of hadrons, which are produced in the different intervals of rapidity.

\section{Short description of the detector and the efficiency of particle registration}

\begin{figure}[htpb]
\centering
  \includegraphics[width=8.0cm, angle=0]{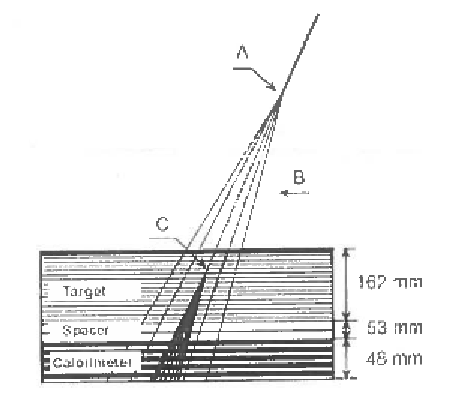}
  \caption{The scheme of detector: A-the collision point in the atmosphere, B-the produced particle tracks, C-hadron interaction in the target.}
  \label{detector}
\end{figure}

The balloon flight took place on the attitudes around 30 kilometers for 160 hours.
The size of detector was 400mmX500mm and 260mm high. Detector consisted of three cameras \cite{lastpub}: upper target block, spacer, and calorimeter, see figure~\ref{detector}.
 
\begin{figure}[htpb]
\centering
  \includegraphics[width=8.0cm, angle=0]{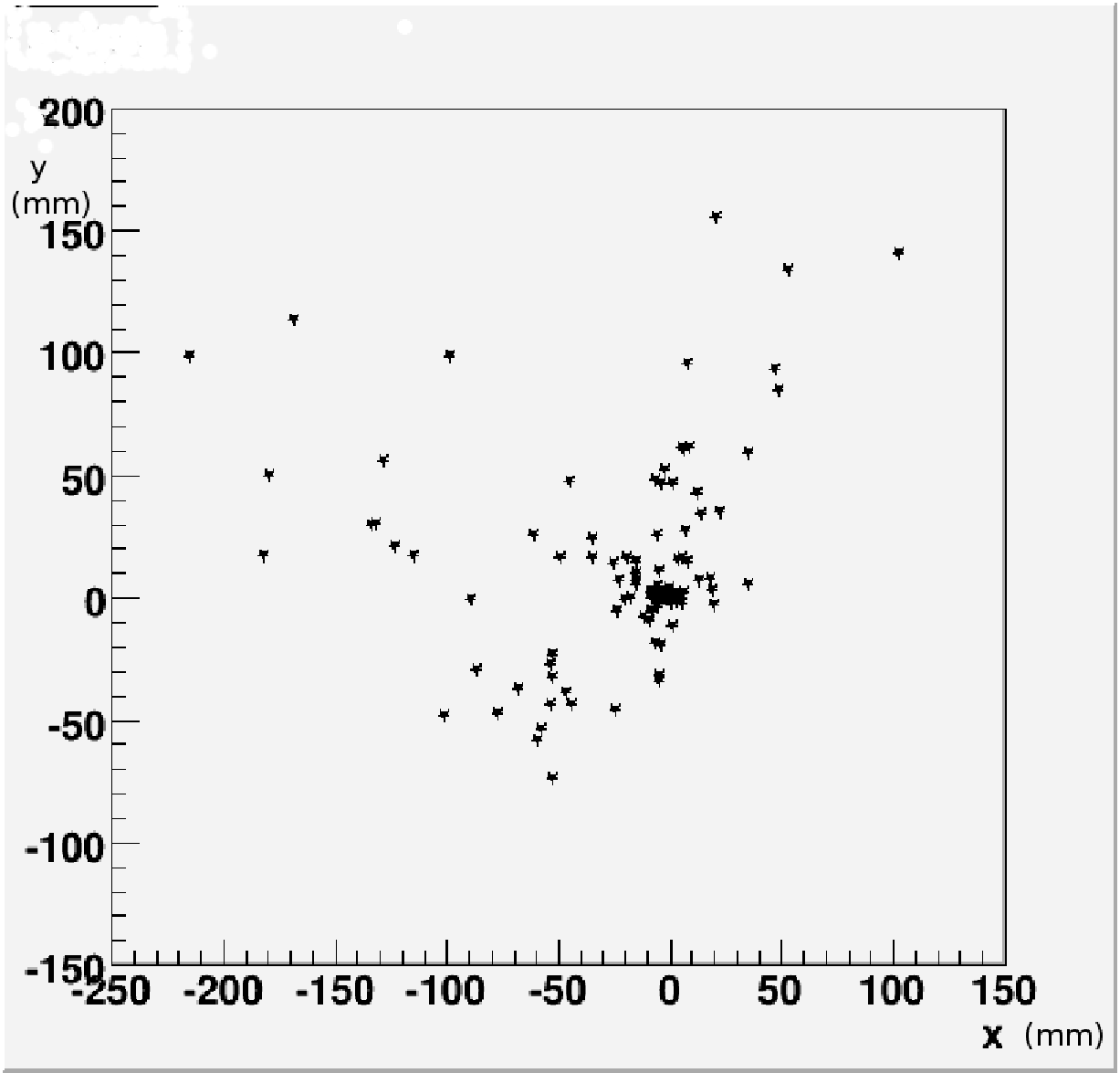}
  \caption{The target diagram of the registered tracks.}
  \label{familyphoto}
\end{figure}

The total thickness of matter corresponds to 0,26 of $\lambda_{freepass}$ that is the free pass without hadron interaction. 
The efficiency of hadron detection is of order 40 percent. 
The electron-photon cascades from secondary hadrons have been developed in the lead layers of the calorimeter and detected as the dark spots in the x-ray films. 
The spot with certain darkness gives us the energy of the particle produced in the interaction. 
A more detailed description of the experimental circumstances can be found in the old publications \cite {stratosphere,lastpub}.
Unfortunately, almost 30 percent of particle tracks have been lost due to the getting out of the scope of the calorimeter.
The positions of 106 observed dark spots are shown in figure~\ref{familyphoto}.

\section{Rapidity distribution: the particle production at the central rapidities and in the forward fragmentation region}

\begin{figure}[htpb]
\centering
  \includegraphics[width=8.0cm, angle=0]{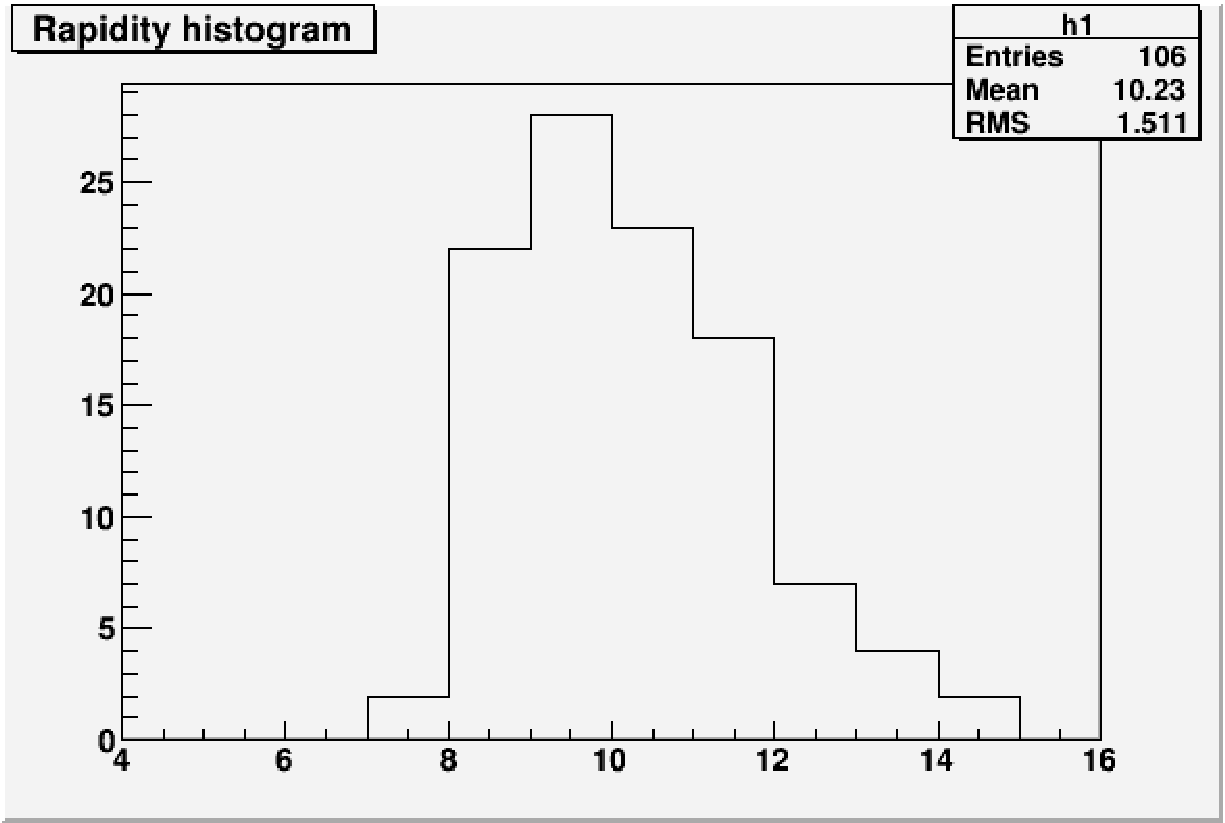}
  \caption{Rapidity distribution of particle tracks in the laboratory system.}
  \label{Yhysto}
\end{figure}

The primary data that we can extract from the table of coordinates and energies of visible tracks are transverse masses
and rapidities, $Y = - 0.5*\ln{((E-Pl)/(E+Pl))} = - \ln{M_t/(2E)}$, where $M_t= \sqrt{Pt^2+M_0^2}$, E, Pl and Pt are energy and momenta of particle and $M_0$ - its mass.
The transverse masses, $M_t$, have been calculated from energy, coordinates of particle track and the distance from the collision point.
Our event was detected in the laboratory system, where all rapidities are positive, see figure~\ref{Yhysto}.

\begin{figure}[htpb]
\centering
  \includegraphics[width=8.0cm, angle=0]{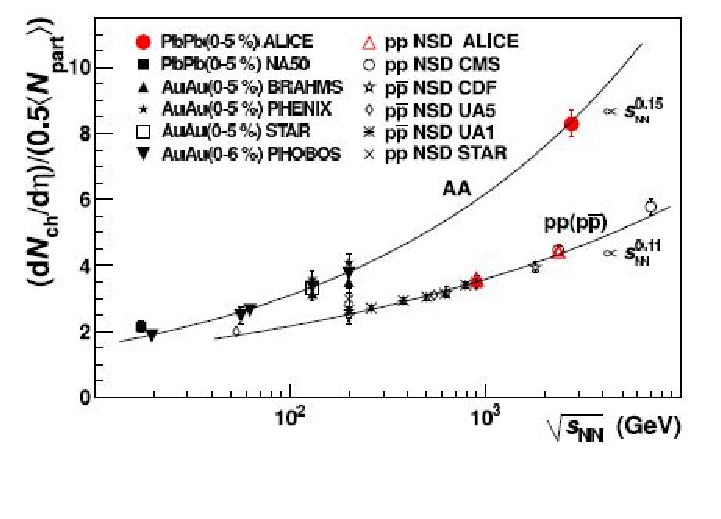}
  \caption{Charged particle pseudorapidity density per participant proton pair for central nucleus-nucleus and non-single diffractive p-p (p-antip) collisions as a function of energy, $\sqrt{s_{NN}}$ \protect\cite{alice}.}
  \label{alice}
\end{figure}

As it was compared with collider distribution, the point near $Y_{lab}$=7.5 corresponds to the center of rapidity spectrum in the center-of-mass system accepted for the experiments on colliders. Two regions of this distribution are interesting from the point of view of contemporary hadroproduction physics: central part of histogram gives us the density of produced hadrons, $dN^h/dY$ at $Y_{cms}=0$, and the area of forward hadroproduction helps to estimate the maximal rapidity of secondary protons $Y^p_{max}$, which depends on the energy of collision per one proton \cite{baryonasymmetry}. The density 
of particles near $Y_{cms}=0$ shows how many protons are collided that gives us the atomic number of a projectile nucleus, see 
figure~\ref{alice}

Unfortunately, the lack of tracks, which are targeting beyond the scope of the detector, is seen in central bins of the histogram at $Y_{cms}=0$.
Nevertheless, we can conclude that $dN^h/dY_{cms}(0) \ge 28$. In addition, we estimate the energy, $\sqrt{s}$, for proton-proton collision that can be calculated from $Y^p_{max}=ln(2\sqrt{s}/M_p)$ = 6.5. The energy per one proton equals to 400 GeV approximately.
In such a way, we should have the hadron multiplicity per a pair of interacting protons of the order 5.0, as it is compared with the LHC plot 
\cite{alice} for A-A collisions, see figure~\ref{alice}. It means that the projectile nucleus, A=2*(28/5.), was at least boron (10 nucleons) or, preferably, carbon (12 nucleons).

Now let us turn to the forward region of rapidity distribution. This is the area of the very-dense dark blot on the target diagram, where definitely no tracks are lost, see figure~\ref{familyphoto}. As will be seen in the next section, particles 
in the central bulky blot are mostly photons or electrons. The hadron contribution appears at lower rapidity $Y_{lab}^p=14$, which corresponds to the maximal proton rapidity in c.m.s. $Y^p_{max}=6.5$. The detailed discussion of proton content of forward 
fragmentation area will be done in the next sections.

\section{Transverse mass distributions}

Actually, in this experiment, we cannot distinguish a type and a charge of particle, but their transverse masses, $M_t$, can be obtained with the following equation:
$M_t = \sqrt{M_0^2+P_t^2}$.

\begin{figure}[htpb]
\centering
  \includegraphics[width=8.0cm, angle=0]{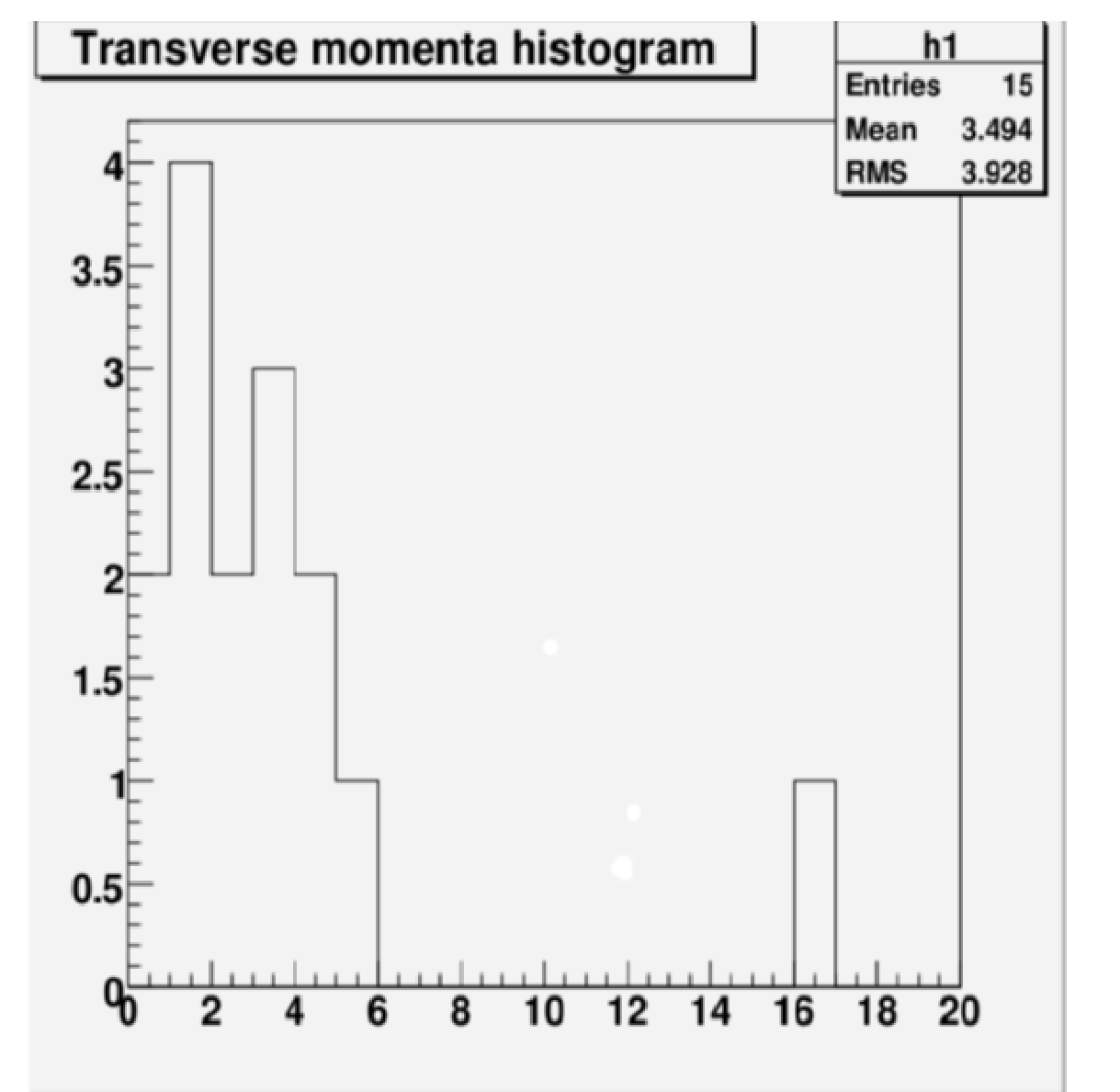}
  \caption{Transverse masse distribution in central rapidity area, $Y_{cms} \le 2$.}
\label{Mthysto_a}
\end{figure}

\begin{figure}[htpb]
\centering
  \includegraphics[width=8.0cm, angle=0]{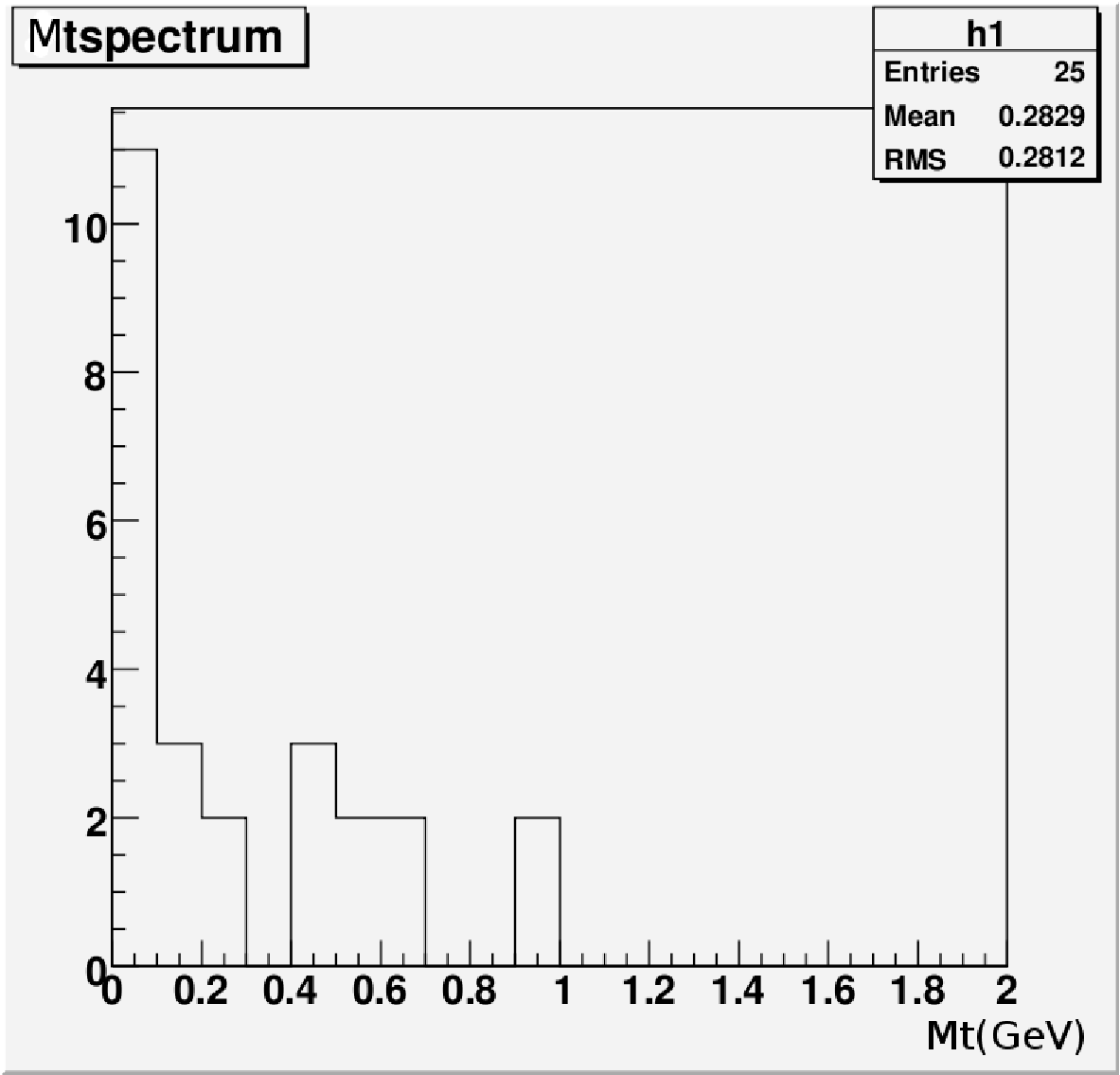}
  \caption{Transverse mass distribution in the region of forward fragmentation region, $Y_{cms} \ge Y_{max}-3$.}
  \label{Mthysto_b}
\end{figure}

The partial histograms show the transverse mass distribution split between a) central rapidity area figure~\ref{Mthysto_a} and b) forward fragmentation region figure~\ref{Mthysto_b}.

\begin{figure}[htpb]
\centering
  \includegraphics[width=8.0cm, angle=0]{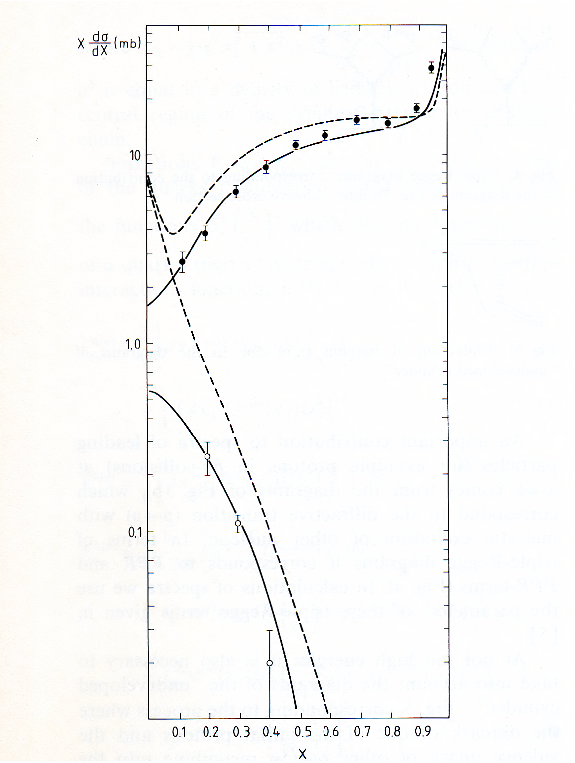}
  \caption{The QGSM description of proton spectrum in the whole kinematical range for p-p collisions at colliders: the peak 
at $x_F$ close to 1 appears due to diffractive dissociation of beam proton.}
  \label{kaidalov}
\end{figure}

The transverse masses in the region of forward particle production, figure~\ref{Mthysto_b} , are small. 
Only two forward tracks have masses near 1 GeV, which is the signature of proton from three pomeron peak that naturally exists 
at the end of inclusive spectra of protons\cite{kaidalov}
, see figure~\ref{kaidalov}. The central rapidity spectrum of transverse masses in figure~\ref{Mthysto_a} shows $M_t$'s in the range from 0 to 6 GeV that corresponds to typical masses of hadrons, excepting one track with the transverse mass 16 GeV. 
This track may be a sign of new physics.

\section{The important differences between astroparticle interaction with the atmosphere and nucleus-nucleus collision at LHC}

On the basis of analyisis discussed above, we have noticed three phenomena that are not expected within the long-time studies of hadroproduction at LHC.
The explanations of these facts are suggested in the framework of the baryonic Dark Matter hypothesis \cite{torusasDM}.

\subsection{Multiplicities in the three Pomeron peak and at the central rapidity area} 

The densities of particle multiplicity that have been detected in the forward and in the central rapidity regions disagree with the pattern of nucleus-nucleus collisions.  The density of produced particles in central rapidity area reports at least  about carbon-nucleus interaction see figure~\ref{Yhysto}, while the presence of only two protons 
in the three Pomeron peak tells us that the projectile was helium. 

\subsection{The circles in rapidity distribution}

\begin{figure}[htpb]
\centering
  \includegraphics[width=8.0cm, angle=0]{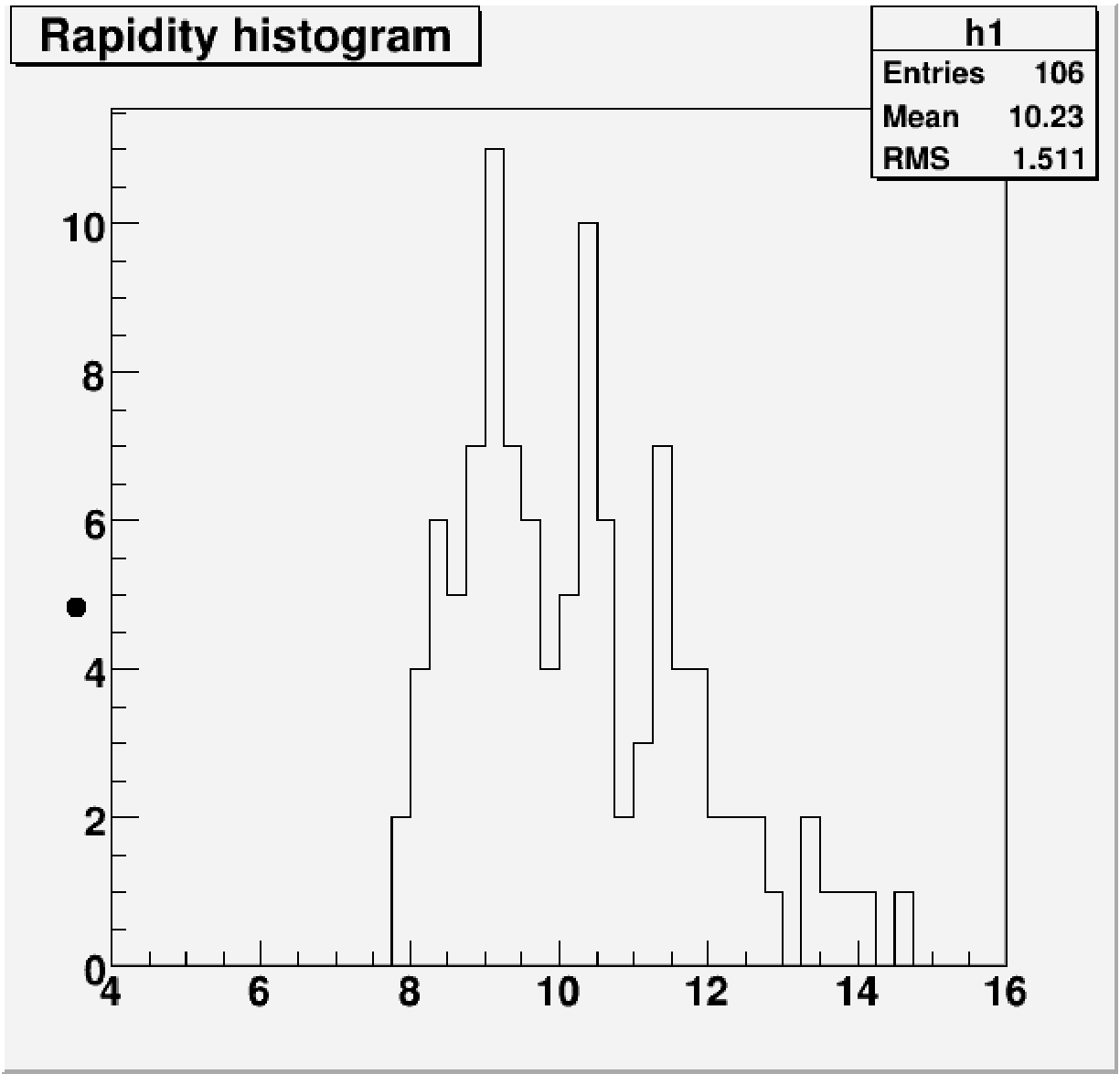}
  \caption{The rapidity histogram with small binning (dY=0.25) shows peaks at different rapidity values that were previously explained as an analog of Cherenkov radiation circles in the nuclear environment \protect\cite{dremin}.}
  \label{Yhystobin025}
\end{figure}

The rapidity distribution with a smaller binning, see figure~\ref{Yhystobin025}, shows the rapidity peaks that means the radiation of hadrons within the circles, which 
were already described by I.M. Dremin \cite{dremin}. New explanation of this and previous phenomena is suggested here. The energy of astroparticle is divided between the components due to the Regge type of structure function that is peculiar for the strongly connected QCD objects  like quarks in proton. Each component gives the proton spectrum with three Pomeron peak at the rapidity,corresponding to its mass. The lightest debris of baryonic DM particle moves with the maximal rapidity and gives the smallest number of nucleons at the forwarding end of spectra. 

\subsection{The detected hadron with the transverse mass 16 GeV.}

\begin{figure}[htpb]
\centering
  \includegraphics[width=8.0cm, angle=0]{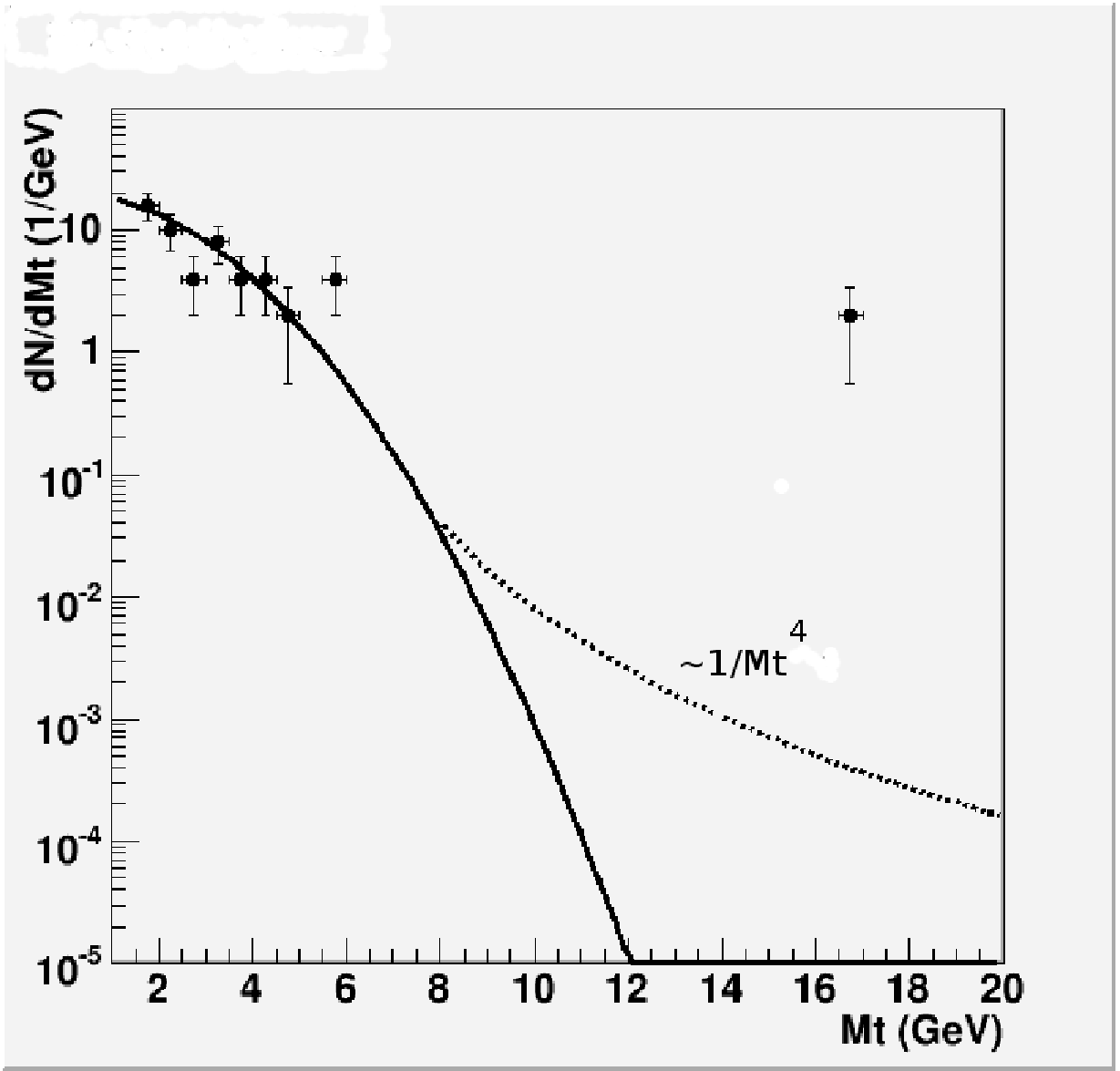}
  \caption{Transverse mass distribution in the central rapidity area and the theoretical fit.}
  \label{Mthystofit}
\end{figure}

The particle was detected with $M_t$=16 GeV that is more than any known hadron mass. The appearence of such particle cannot be explained by gluon-gluon fusion, see figure~\ref{Mthystofit}.

It is natural to expect also a second particle with a similar high transverse mass, which was obviously lost beyond the scope of detector.
The theoretical expectation for the jets from gluon fusion is proportional to $1/(Pt^4)$ and  gives us 0.0004 particles in this region that disagrees with the observation. It meens that particle production at $ M_t$=16 GeV can not be just fluctuation of transverse momenta of produced hadrons.

\subsection{The suggestion of baryonic DM decay}

All these signatures may declare itself the interaction of new heavy QCD particle.  
This baryonic DM particle, as it was recently suggested \cite{ICHEP, mass}, can have heavy masses and decay at the collision on
two similar DM particles with lower masses close to 14 GeV and few nucleons, which produce multiple particles, see Appendix.  The spectra of baryons in this shower demonstrate the expected peaks in the forwarding  rapidity region.
The production of baryonic DM particles on colliders is the process of very small probability \cite{torusasDM} and these QCD objects decay immediately. They  certaily are to be realized only at the processes in the centers of galaxies , so was named "astroparticles". 

\section{Appendix. Hypothesis of hadron mass progression}

The dependence of average transverse momenta on the mass
has been studied recently in \cite{ICHEP,mass}. If we imagine a
symmetric point between meson and baryon masses of a given quark flavor, the mass distance between points of one hadron generation and the next one can be estimated with the factor 2,71828. It means that we have the geometric progression for the masses of hypothetical hadrons,
as in equation~\ref{massprog}.

begin{equation}
M_n=0.251*e^{n-1}, (GeV)
\label{massprog}
end{equation}

\begin{figure}[htpb]
\centering
  \includegraphics[width=7.0cm, angle=0]{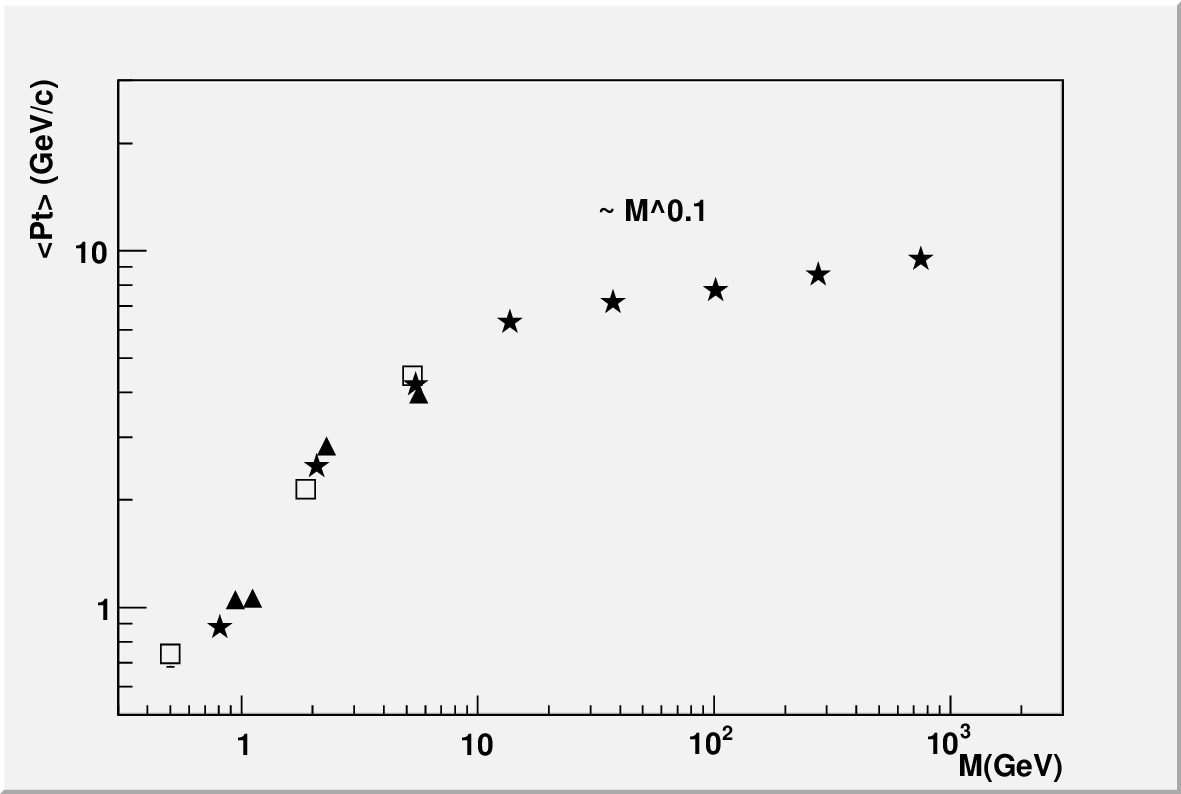}
  \caption{Average transverse momenta vs. masses of hadron generations: mesons - empty squares, baryons -
black triangles. The masses of hypothetical neutral heavy hadron states are shown with the black stars. The hadron mass
progresion above the beauty hadron masses is extrapolated with average pt dependence, where average  pt  is proportional to $M^{0.1}$.}
  \label{ptvsM}
\end{figure}

This equation describes the distances between existing generations of hadrons: 0.251, 0.68, 1.86, 5.04. 
The extension of this sequence provides the new neutral heavy hadron states with the following
masses: 13.7, 37.3, 101.5, 276, 750 GeV, and so forth, see figure~\ref{ptvsM}.

These hypothetical hadrons represent heavy neutral hadron states. From this point of view, the recently observed
pentaquarks and tetraquarks are the debris of invisiblel neutral heavy hadron states.

\section{Conclusions}

The analysis of hadroproduction event in the stratosphere has been carried out in the framework of modern conventions about high energy nucleon-nucleon interactions in colliders. The inclusive distributions have been compared with the spectra that are typical for LHC data and approved with the Quark-Gluon Model.
The conclusions are the following:

1) The rapidity distribution tends to have a constant density of charged particles per unit of rapidity in the central region ($Y_{cms}$=0). The value of central multiplicity corresponds to carbon nuclei (12 nucleons) collision with the CNO nucleus of the atmosphere. The target fragmentation part of spectra is not seen due to the small size of the detector. The maximal rapidity corresponds to the energy per nucleon of the order of 400 GeV in the center-of-mass system that gives the energy near 5 TeV for a carbon-carbon collision.

2) There are particles of small masses: photons, electrons, muons, and pions in the very forward region of rapidities. Two particles have transverse mass near 1 GeV and are interpreted as protons that contribute into the visible peak at the end of rapidity spectrum. The three Pomeron peak is the expectable feature of proton spectra in proton-proton collision \cite{kaidalov}. Two protons in this peak would report a helium projectile collision in the atmosphere. But the detailed research makes us assuming that there were new type of hadron with strong connections between components. Such composite projectile tells us about some massive QCD structure decaying into two less massive particles plus few nuclei. 

3) One track with $M_t$=16 GeV shows the outstanding value of transverse mass that is far from the mass range of all other well known hadrons. The theoretical expectation for gluon-gluon fusion, $1/P_t^4$ gives 0.0004 particles in this region, 
This makes us concluded that a new state of hadron matter can exist at masses of order 14 GeV. Such heavy neutral hadronic states has been pronounced in the recent papers \cite{ICHEP,averagept,torusasDM} as the neutral multi baryon structures with the hidden baryon charge, which are good candidates for Dark Matter.

4) After the discourse on the specifics of stratosphere x-ray detection, we propose the similar high-altitude cosmic ray experiment that can be useful, on one hand, as a good supplement to the LHC measurements. On the other hand, they are able to discover events of first collision of unknown  astroparticles  in the full kinematical range, while colliders are studying nuclear interactions only in the central rapidity area. 
The modern experiment on the basis of large calorimeter with optical fiber detector was already planned on the mountain altitudes 
\cite{stranaproject} in order to search for the primary nuclear collisions of TeV energies and higher. Such experiments should be continued, but are prefered to be constructed with the application of up-to-date electronic methods. 

\section{Acknowledgements} 
The experiment discussed in this article was carried out owing to the enthusiasm and efforts of Konstantine Aleksandrovich Kotelnikov.
The authors express their gratitude to Prof. O. Kancheli and Dr. K. Boreskov (ITEP) for multiple theoretical remarks and the committed focus on our research. This paper has been written before December of 2019, when K. A. Kotelnikov passed away.

\end{document}